# Transition from a ferromagnetic insulating to a ferromagnetic metallic state in nanoparticles of $Nd_{0.8}Sr_{0.2}MnO_3$ : Study of the electronic - and magneto - transport properties

S. Kundu[a] and T. K. Nath[*]

*Department of Physics and Meteorology, Indian Institute of Technology, Kharagpur, West Bengal 721302, India*
[a]*email: souravphy@gmail.com*

***Abstract***

*A detailed investigation of the electronic - and magneto - transport properties of $Nd_{0.8}Sr_{0.2}MnO_3$ with the variation of grain size (down to 42 nm) is presented here. Interestingly, we observe that the ferromagnetic insulating state is suppressed and a metallic state is stabilized as the grain size of the sample is reduced. As a result, metal insulator transition is observed in this low doped manganite which is insulating in nature in its bulk form. Destabilization of polaronic order in the ferromagnetic insulating state due to enhanced surface disorder on grain size reduction has been attributed to this effect. A phenomenological model has been proposed to represent the concept of destabilization of polaron formation in the surface region of the nano grains. Resistivity and magnetoresistance data have been carefully analyzed employing different suitable models. Electrical third harmonic resistance has been measured to directly probe the electrical nonlinearity in the samples.*

---

[*]Corresponding author: tnath@phy.iitkgp.ernet.in

Tel: +91-3222-283862
Area code: 721302,
INDIA

## 1. Introduction

The manganites showing a rich variety and many unusual aspects in their electronic and magnetic properties have been the focus of research interest since last few decades [1]. The interest in these materials has mainly grown up because of the discovery of colossal magnetoresistance during 1990s. Even after a detailed study carried out on such materials, many of their properties have not been properly understood. One such fascinating case is the ferromagnetic insulating state of manganites. We have addressed in this paper the interesting issue of grain size effect on the ferromagnetic insulating (FI) state of the manganite $Nd_{0.8}Sr_{0.2}MnO_3$. Effect of reduction of physical dimension or the effect of nanosize grains is found to have many fascinating effects on manganites which are intrinsically complicated systems having a close interplay between their spin, charge and lattice degrees of freedom [2, 3]. However, most of the studies regarding size effect are carried out on the optimally doped [4,5] or on the half doped charge ordered manganites [6,7]. Few works on nanometric effect have also been carried out on the over doped side of manganites [8,9]. To the best of our knowledge, a detailed study of the effect of reduction of the grain size on the FI regime of a manganite has not been carried out in detail so far.

The most interesting feature about the ferromagnetic insulating state which is generally observed in the low doped regime of manganites is that it cannot be understood in the framework of double exchange. A strong electron - phonon coupling is believed to have a key role for the formation of such a FI state. Carriers are localized due to this strong coupling with the crystal lattice and drive the system insulating. Literally, the carriers localized to the lattice are called polarons [10]. Several different origins of this carrier localization including J-T type distortion, Coulomb potential fluctuation due to the presence of trivalent and divalent ions, spin disorder etc. have been discussed in details till date. Most importantly, under such localization, evidently, the transport of electron becomes predominantly phonon assisted i.e. with the increase of temperature conductivity of the electron increases. The system behaves like insulator/semiconductor ($d\rho/dT<0$). Efforts have been made to describe the electronic properties in such a state of manganites in terms of polaronic transport, namely, adiabatic or non - adiabatic small polaron hopping (SPH), variable range hopping (VRH) etc [11,12]. Colossal magnetoresistance effect [13] as well as a large electroresistance [14] has been found in the FI state of the manganites. All these transport related effects are very useful for technological applications. So, it appears that the ferromagnetic insulating state of manganites is very important from physics point of view as well as having potentials for applications. In that case a detailed study of the electronic - and magneto - transport properties of a ferromagnetic insulating manganite seems very plausible and, moreover, the study of the effect of grain size will obviously enrich the investigations. We have carefully studied the electronic - and magneto- transport properties of the manganite $Nd_{0.8}Sr_{0.2}MnO_3$ with the variation of grain size down to the nanometric regime. The most fascinating observation of this study is that the FI state is destabilized and a metallic state emerges when the grain size of the system is reduced. The general notion about the nano size effect is the effect of enhanced grain boundary which hinders the electronic transport across the grains and increases the resistivity of the system [5]. Exceptions have been found for most of the half doped charge ordered insulating systems where reduction of grain size renders the system metallic due to the destabilization of the ordering phenomena [6,7]. However, our study is new of its kind, and reveals that even in the low doped ferromagnetic insulating regime, metallic behaviour can be observed as an outcome of reduction of the grain size.

## 2. Experimental Details

The polycrystalline samples were synthesized through chemical pyrophoric reaction route. The starting materials were high purity (99.9%) $Nd_2O_3$, $Sr(NO_3)_2$, $Mn(CH_3COOH)_2$ and Triethanolamine (TEA). The details of preparation process can be found elsewhere [5]. The as prepared powder was divided into four parts and sintered at different temperatures (namely, 750 ºC, 850 ºC, 950 ºC and 1150 ºC) in air for five hours. The objective was to produce samples of different grain size with the same stoichiometry. The finally obtained samples are palletized and used for different experiments. Throughout this paper the samples are designated as S750, S850, S950 and S1150, respectively, after their corresponding sintering temperature.

The structural characterization has been carried out through high resolution x - ray diffraction (HRXRD), Field Emission Scanning Electron Microscopy (FESEM) and High ResolutionTransmission Electron Microscopy (HRTEM). Electronic- and magneto- transport properties have been studied employing a closed cycle helium refrigeration cryostat fitted in a superconducting magnet (8 Tesla field) from Cryogenics Ltd. U.K., down to a temperature of 2 K. The magnetic properties have been measured employing a homemade ac susceptometer coil fitted inside the helium cryostat and a homemade vibrating sample magnetometer down to 77 K. Both these instruments employ a high precision Lock-In-Amplifier (model-SR 830) for signal detection and Lakeshore made high precision PID temperature controller (model-331S) for controlling the temperature.

## 3. Results and Discussion

The high resolution x - ray diffraction study reveals that all the samples are of single phase with no detectable impurity peaks. To have a detailed structural information, Rietveld refinement of the x - ray spectra has been carried out with the aid of Maud programme. It is found that all the samples crystallize in orthorhombic structure with *Pnma* space group. A typical refinement data for S750 sample is shown in Fig. 1 (a). This displays a reasonably good quality fitting of the experimental data. The obtained informations from refinement are summarized in the Table 1. One can observe that there is no significant and systematic variation of lattice parameters with the change in grain size (sintering temperature). The HRTEM image of S750 (Fig. 1 (b)) and size distribution of grains (Fig. 1(c)) indicate that the average grain size of this sample of S750 is about 42 nm. The SAD pattern (Fig. (d)) and high resolution lattice image (Fig. 1 (e)) reveal poly crystalline nature and good crystallinity of S750 sample, respectively. The FESEM images of S850 and S950 samples display that the grain size of them are nearly hundred and few hundreds of nanometer (Fig. 2(a) and (b)), respectively. The S1150 sample is having a micrometer order grain size (Fig. 2 (c)) and expected to have a bulk like property. The calculated chemical formula of S750 sample from its EDS spectra as shown in Fig. 2 (d) is $Nd_{0.77}Sr_{0.23}Mn_{0.98}O_{3.00}$. This confirms that the change in the grain size (or sintering temperature) does not affect the stoichiometry of our samples to a considerable extent.

## I. Electronic - transport

The measured resistivity of the samples as a function of temperature and under different magnetic fields, employing a conventional four probe technique, is shown in the Fig. 3. Interestingly, except S1150, all the other samples show a metal - insulator transition (MIT). The insulating behaviour of S1150 (Fig. 3(a)) is in accord with the reported phase diagram of bulk $Nd_{0.8}Sr_{0.2}MnO_3$ [1]. Even under high magnetic field (8 T) it does not show any metallic behaviour. However, with a slight reduction of sintering temperature or grain size, insulating behaviour is suppressed and metallicity ($d\rho/dT > 0$) is observed (namely, for S950) as shown in Fig. 3(b). The MIT temperature gradually increases with the decrease of grain size indicating an enhanced stabilization of the metallic state over a wide temperature regime. All the samples showing MIT also display an upturn of resistivity in the low temperature regime. This kind of resistivity minima have been observed in many polycrystalline manganites earlier. Nevertheless, such a systematic evolution of resistive behaviour of the samples indicates a clear correlation between grain size and suppression of insulating state to a metallic one. Our EDS data confirms that the observed behaviour is not due to the change of stoichiometry which may occur as a result of variation of sintering temperature. Moreover, a change of stoichiometry equivalent to a change of $Sr^{2+}$ doping level from 20% to nearly 30% is required to observe any metallic state in Nd-Sr-Mn-O system. Such an enormous amount of change of stoichiometry is most unlikely to occur in this system due to the change of sintering temperature. Hence, we attribute this novel observation of emergence of metallic state in the insulating regime of the low doped manganite to size reduction of the system down to the nanometer scale (42 nm).

A thorough analysis of the resistivity data has been carried out to understand the electronic - transport process in the samples. The S1150 sample is insulating over the whole temperature range. However, a sharp increase in resistivity takes place just below 100 K (Fig. 3(a)). Resistivity increases up to a few orders of magnitude in the lowest measured temperature. A small change in the slope of resistivity vs. temperature data in the low temperature regime is also observable just around 35 K. One can easily notice that in the ac susceptibility vs. temperature data of S1150 sample there are two magnetic transitions (except the ferromagnetic - paramagnetic transition around 140 K), around 90 K and 35 K, respectively (Fig. 4). Clearly, the magnetic behaviour correlates well to the electronic - transport of S1150. Actually, the S1150 sample shows a cluster - glass like behaviour below 100 K, as previously explored by us [15]. So, most likely the change in the magnetic state has a role to play in sharp change in resistivity below 100 K. However, we have analyzed our data employing different models of electronic transport to have a more insight in the mechanism. Firstly, the small polaron hopping model [12, 16] of the form $\rho = \rho_0 T \exp(W/k_B T)$, where W is the activation energy, has been employed. This equation is found to fit well to the experimental resistivity data of S1150. To calculate the relevant parameters, ln(ρ/T) is plotted as a function of 1/T. The plot shows straight lines with different slopes at different temperature regime as shown in Fig. 5 (a). This is indicative of the fact that a single set of parameters associated with the SPH model is not enough to describe the electronic transport of S1150 over the whole temperature range. The obtained fitting parameters for S1150 for the different temperature regime (including the paramagnetic insulating regime) are summarized in the Table 1. In Table 1 the calculated activation energies are listed along with the corresponding temperature regime for S1150 sample. Most interestingly, one can observe that the sharp increase in resistivity of S1150 in the low temperature regime is not due to the increase in the activation energy with the decrease of temperature. In fact, activation energy decreases with the decrease of temperature according to the small polaron hopping model. It is the increase of the pre - exponential factor ($\rho_0$) which drives the resistivity to an extremely high value in the low temperature regime. The term $\rho_0$ is expressed as $\rho_0 = [k_B/\nu_{ph} N e^2 R^2 C(1-C)] \exp(2\alpha R)$, where, $\nu_{ph}$ is optical phonon frequency, N is the number of metal ions per unit volume of the sample, $R$ is average hopping distance, C is fraction of sites occupied by polarons and α is known as tunnelling probability or the electronic wave function decay constant [17]. It is difficult at this stage to conclusively explain which parameter/parameters is/are responsible for the increment of $\rho_0$ in the low temperature regime based on only resistivity data. However, presumably, increase in the value of α with the decrease of temperature is the reason as other terms of $\rho_0$ are more or less constant of the system. Increase in the value of α indicates more localization of the electrons. The enhanced magnetic disorder with the decrease of temperature [15] in S1150 is possibly the reason for this enhanced localization of electrons.

It is a well known fact that transport in the paramagnetic insulating regime of manganites is also governed by polaron formation [18]. We have also employed the SPH model to describe the transport in the paramagnetic regime of S950, S850 and S750 (Fig. 5(a)) samples, successfully. The experimental data could be fitted well with the model and the obtained parameters are given in the Table1. Interestingly, it is observed that the activation energy increases with the decrease of grain size. A similar trend with similar values of the activation energy was found in the system $Nd_{0.7}Sr_{0.3}MnO_3$ [19]. Efforts have been made to describe the electronic transport by employing another frequently used model, namely, Mott's Variable Range Hopping Model (VRH) of the form $\rho = \rho_\infty \exp[(T_0/T)^{1/4}]$ [ 20] in the paramagnetic insulating regime of the samples showing MIT. In this case also satisfactory fitting has been done with the obtained values of density of states as summarized in the Table 1. The values of the density of carriers at the Fermi level ($N(E_F)$) have been calculated from the relation $T_0 = 16\alpha'^3/k_B N(E_F)$, assuming $\alpha' = 2.22$ $nm^{-1}$[17]. It is found that $N(E_F)$ is small in case of S750 and S850 compared to that of S950. It has also been observed from the obtained $R^2$ values of least square fitting that VRH model fits better compared to SPH model in case of S750 and S850, whereas, it is just the reverse situation in case of S950.

The metallic regime of the samples showing a MIT has been well explained assuming two dominant scattering phenomena, namely, electron - electron and electron - magnon scattering, respectively. It is known that the electron-electron scattering has a $T^2$ dependence, whereas, the second order electron- magnon scattering goes as $T^{4.5}$ with temperature. The total resistivity of the form $\rho = \rho_0 + \rho_2 T^2 + \rho_{4.5} T^{4.5}$ [21,22] was fitted to the experimental ρ-T data in the metallic regime (Fig. 5(b)) and the obtained coefficients are listed in the Table 2. We observe that $\rho_2$ for all the samples has a very high value compared to the reported values of some other manganites systems [23], apparently reminiscent of a strong electron - electron (e - e) scattering. The obtained coefficient $\rho_2$ decreases with the decrease of grain size. The coefficient $\rho_{4.5}$ is negligible for S950 sample but have reasonably high values for S750 and S850 samples.

The observed low temperature resistivity minima and resistivity upturn for all the samples with a MIT has been focussed and carefully analyzed. Observation of resistivity minima in many polycrystalline systems has been an intensely discussed issue for a long time. Possibility of a Kondo effect or competition between inelastic scattering processes with the e - e interaction in the low temperature regime was mentioned by many groups [24, 25]. However, Ll. Balcell *et al.* [26] experimentally showed the effect of Coulomb blockade of electrostatic origin as an operating mechanism behind the increase of resistivity in the low temperature regime in granular manganites with grain size less than 0.1 μm. The electron requires an effective charging energy ($E_C$) to be separated from a positively charged grain and this energy equals $e^2/4\pi\varepsilon d$ where d is the average grain size (assuming spherical grain) and $\varepsilon$ is the permittivity of the material. The empirical equation $\rho = \rho_0 \exp(\sqrt{\Delta/T})$, where $\Delta \approx E_C$, is found to hold quite well below the minimum temperature ($T_{min}$) of our samples showing a MIT. A typical fitted data for S750 sample at 0 T field is shown in the inset of Fig. 5 (b). The obtained values shows that $E_C$ increases by two orders of magnitude in the nanometric S750 and S850 samples compared to S950 sample (Table 2). This is indicative of the fact that the effect of Coulomb blockade is only dominant for samples with smaller grain sizes (few tens of nanometer) as shown in Ref. 22. We have analyzed the low temperature resistivity of S750 sample in little more detail. The minima position ($T_{min}$) and the depth of the minima $[(\rho(7K)-\rho(T_{min})/\rho(7K)]$ is plotted as a function of applied magnetic field as shown in the inset of Fig. 5 (c). Very, small dependence of both the quantities on magnetic field is observed. These rule out the possibility of any Kondo-like effect in the sample. However, to describe the low temperature field dependence of the resistivity upturn precisely, one has to include other field dependence scattering processes and barriers especially in case of manganites. In this context, we want to mention that on application of magnetic field, an increase in $E_C$ of S750, precisely, 7.8 K at 0 T to 13 K at 8 T, is observed. Having a completely electrostatic origin, $E_C$, in principle, should not have any magnetic field dependence. So, it is an indication that Coulomb blockade is not the only contribution to the low temperature resistivity minima in the sample under consideration.

## II. Magneto - transport

An elaborated study of the magneto - transport properties of the samples has been carried out with the variation of temperature as well as magnetic field. The temperature variation of magnetoresistance (MR% = $([\rho(H)-\rho(0)]/\rho(0)\times100)]$ under 8 T magnetic field for all the samples is shown in the insets of Fig. 3. The nature of variation shows that negative MR increases with temperature, attains a very high maximum value (80% - 90%) and again decreases as the paramagnetic state (refer to Fig. 4 for magnetic behaviour) is approached for all the samples. For the samples showing MIT (S750, S850 and S950) the maxima in MR is expectedly observed around the corresponding MIT temperature due to the enhanced phase instability and phase separation tendencies around a metal insulator transition. Interestingly, S1150 sample having no MIT display a maxima around the low temperature regime (~ 60K). We can recall that this is the temperature regime where the resistivity increases sharply. Magnetoresistance of the samples has also been measured as a function of magnetic field at different fixed temperatures and shown in the Fig. 6. All the samples show a qualitatively similar behaviour in the field dependence of the MR. The negative MR in the ferromagnetic regime of the samples gradually increases with the increase of magnetic field and show a non - saturating nature even upto 8 T field. To understand the mechanism of the observed magnetoresistance a number of models have been employed to fit our MR vs. H data. Firstly, a general scaling law of the form $MR = -\beta H^{\gamma}$ [27] is employed to fit our experimental data. This model is generally used to describe the magneto - transport in spin - glasses. Keeping in mind the observed glassy nature of our system [15] we have employed this equation on our samples. We have observed that this equation fits well to the MR vs. H data of our samples in the low temperature regime (<50 K) only, and deviates from the measured value of MR as the temperature is increased (not shown). It is found that $\gamma$ lies in the range of 0.4 - 0.7 for all the samples in the above mentioned low temperature regime. For canonical spin - glasses $\gamma$ is found to be around 2. Another proposed mechanism for magnetotransport in manganites, namely spin dependent hopping [28], is also employed to describe the observed nature of MR vs. H data. According to this model an electron undergoes spin dependent hopping between nearby magnetic clusters and MR is given by F(T)$B$(x) in the ferromagnetic regime and P(T)$B^2$(x) in the paramagnetic regime, where, F(T) and P(T) are temperature dependent constants, $B$ is the Brillouin function and $x=g\mu_B J(T)H/k_B T$. We have on the other hand, employed an equation of the type MR = F(T)$B$(x) +P(T)$B^2$(x) keeping in mind the phase separation tendencies of manganites and obtained a satisfactory fitting (Fig. 7(a)) with the experimental data at all temperatures for all the samples. Unfortunately, at many temperatures the obtained parameters are not meaningful though the fitting seems to be quite well. The typical values of magnetic cluster size (calculated from the obtained J values) in terms of lattice units (l.u) and that of ferromagnetic and paramagnetic phase fractions (F% and P%) obtained from the fitting parameters for S1150 and S950 at different temperatures are shown in the Fig. 7 (b), (c), (d) and (e). The average order of magnitude of the cluster size, the nature of temperature variation of phase fractions seem physically meaningful and conform to the already reported values of those quantities in manganites [29]. However, at some particular temperatures the value of the cluster size is very small and unphysical as evident from Fig. 7 (c) and (e). The obtained parameters are even more unphysical in case of nanometric S850 and S750 samples and not reported in this paper. Finally, a more frequently used model, found to hold good in many double exchange systems, proposed by Raychowdhury *et al.*[30], on the other hand, describes the magnetotransport very well, for the nanometric S850 and S750 samples. The basic concept of this model is that in the low field regime spin polarized tunnelling (SPT) of carries occurs between the nearby highly spin polarized grains on field induced alignment of the magnetic moments of them and at higher field regime MR is predominantly due to the intrinsic (INT) double exchange mechanism inside each grains. This model takes into account the pinning of the domain boundaries at the grain boundary and, thus, justifies its application in our nanometric samples (S850 and S750) with enhanced grain surfaces. The fitting of the equation $MR = -\int_0^H [A\exp(-Bk^2) + Ck^2\exp(-Dk^2)]dk$ ('$MR_{SPT}$') $-JH-KH^3$ ('$MR_{INT}$') where, A, B, C, D, J and K are adjustable fitting parameters, to the MR vs. H data is shown in the Fig. 6 (a) and (b). From the fitting parameters we have calculated the spin polarized part $MR_{SPT}$ and the intrinsic part $MR_{INT}$ (at 8 T field) of total MR separately at each temperature and plotted in the Fig. 7(f). Temperature variations of these quantities are similar for both S850 and S750 samples and are physically meaningful. We also notice that intrinsic MR is almost constant and high nearly upto 200 K whereas, spin polarized tunnelling MR decreases sharply just above 100 K to a value of nearly zero.

We, now, discuss the possible origin of the observed metallic behaviour due to the reduction of size and subsequent experimental observations. In this context, we want to mention that the effect of hydrostatic pressure (< 8GPa) can produce a similar effect (metallic behaviour) on low doped manganite in the FI state as reported by S. Arumugam *et al* [31]. A closer view to our crystallographic data reveals that there is no significant change in the lattice parameters and unit cell volume with the change of the grain size (Table 1). So, any kind of pressure, whatever may be the origin, which may operate on the grains, is negligibly small as calculated by employing Birch - Murganhon equation of state [32] which involves the change in unit cell volume to estimate the pressure. So, it appears that, other nanometric effects like enhanced grain surface and related surface effects have a major role to play towards the observation of metallic behaviour. It has been mentioned earlier that the mostly believed concept of formation of FI state in the low doped manganites is the formation of polarons. So it can be conceived immediately that a destabilization of the polaron formation and enhancement of the double exchange mechanism takes place as a consequence of the reduction of the grain size to support a metallic behaviour. The nano size grains can be visualized as a magnetically ordered core with a disordered shell. The boundary of the nano grains consists of numerous crystallographic defects and broken bonds. We believe that a destabilization of the polarons takes place predominantly in the surface of the nano garins. As a consequence, double exchange, may be weak in nature, is operative in the surface region as long as the grain is in the ferromagnetic state producing a metallic behaviour over a certain regime of temperature. If the polarons are orbital in nature as discussed by many groups earlier [33,34], then the situation proposed above is highly probable. The reason is that any kind of ordering phenomena is clearly not possible on the back ground of defect full crystal (surface region). We, hereby, propose a simplified phenomenological picture of nanosize grain as depicted in the Fig. 5(d) keeping in mind that the polaronic order, ideally, should gradually be destabilized from the core to the boundary of the grains. In our picture green color represents strong polaron formation whereas, the yellow colour is to indicate a weak polaronic behaviour. Under these circumstances, an electron will find a low resistive path from one grain to another through the surface with the aid of sufficient overlapping of the grains and can migrate via double exchange. The more insulating core with a stronger polaronic order, on the other hand, has no significant role to play in the transport of electrons. With the decrease of grain size the surface to volume ratio of the grains increases and this is reflected in the increasing stability of the metallic phase over a wide temperature range with the reduction of grain size. However, by no means it can be confirmed whether a complete collapse of the polaronic behaviour takes place at the grain surfaces. Most likely, a competing situation between polaronic order and double exchange persists at the surface and is reflected in the very high value of resistance (order of K Ohm) even in the metallic state of our samples compared to that of $Nd_{0.7}Sr_{0.3}MnO_3$ (order of ohm), a prototype metallic - ferromagnetic double exchange system of the same family [19]. The whole situation can be thought as a surface phase separation of manganites. The surface of the nano grains is combination of ferromagnetic insulator and ferromagnetic metal, whereas, the core is a FI.

Supportive evidences are provided in favour of our phenomenological model. Firstly, we look into our ac susceptibility data. We notice that the susceptibility data of S1150 ($\chi_{ac}$ vs. T plot) in the Fig. 4 show a sharp FM-PM transition at around 140 K indicating a comparatively homogenous magnetic order in it. It also shows another two magnetic transitions with the decrease of temperature at around 90 K and 35 K, respectively. As the grain size is decreased it can be noticed that the FM-PM transition as well as the other two transitions are broadened and gradually smeared out. In case of S750 the FM-PM transition is completely flattened and the low temperature magnetic transitions are no more detectable at all. All these features directly or indirectly indicate an increasing magnetic disorder with the decrease of grain size which may occur as an effect of enhancement of grain surface where a large distribution of magnetic exchange interaction is present. Moreover, the M-H behaviour shows (inset of Fig. 4) that coercivity of the samples in the low temperature regime (80 K) increases with the decrease of grain size. This is precisely the outcome of surface disorder. Large number of defects around the grain boundary provides easy pinning centres for the spins producing a large magnetic anisotropy in the system. Hence, all the magnetic measurements corroborate the concept of gradual increase of surface disorder with the decrease of grain size.

Secondly, we have tried to provide a more direct experimental proof of the destabilization of polarons due to the decrease of grain size in the nanometric regime. To achieve this we measure the electrical nonlinearity in our systems as it was experimentally observed previously that polaronic transports are nonlinear in nature [35,36]. An ac current (sinusoidal of 2μA amplitude and 77.3 Hz frequency) is applied to the sample under study by two probes attached on it and the developed voltage is measured by another two probes placed in between the current probes (linear four probe configuration), employing a lock- in-amplifier (SR830). Any voltage that appears at frequency multiple of the frequency of the excitation current can be regarded as the outcome of an electrical nonlinearity. Observed finite value of 3$^{rd}$ harmonic voltage (voltage measured at 3ω ($V_{3\omega}$), ω is the frequency of the driving ac current) in manganite was previously attributed to nonlinear coupling between electric field and polarons [37]. The measured 3$^{rd}$ harmonic resistivity ($R_{3\omega}= V_{3\omega}/I$) of S1150 and S750 are shown in the inset of Fig.5 (a). Interestingly, $R_{3\omega}$ of S1150 sample shows two prominent peaks and a small peak in the low temperature regime where the resistivity increases sharply, indicating a strong nonlinearity in its resistive properties. On the other hand,measured $R_{3\omega}$ for S750 sample is similar to its ρ vs. T behaviour and does not display any special features. In both the cases the magnitude of $R_{3\omega}$ decreases with the increase of temperature (in the paramagnetic regime). At this point a detailed analysis and interpretation of the observed $R_{3\omega}$ is not possible due to the lack of theoretical and conceptual supports from the literature on this matter. We, only infer from the observed high value of $R_{3\omega}$ in S1150 and its comparative low value (two orders of magnitude) in S750 that the polaronic order, supposed to be the cause of the electrical nonlinearity, is destabilized in the nanometric S750 sample supporting our phenomenological model. Moreover, the striking difference of the nature of nonlinear electronic behaviour directly indicates the significant difference in the intrinsic mechanism of electron transport between the bulk and nanometric manganite of $Nd_{0.8}Sr_{0.2}MnO_3$. Such a difference is difficult to realize on the basis of conventional resistivity measurements.

## 4. Conclusions

In summary, we have carried out a detailed investigation on the grain size effect of low doped ferromagnetic insulating manganite $Nd_{0.8}Sr_{0.2}MnO_3$ on its electronic - and magneto - transport properties. We have observed the fascinating phenomena of metallic behaviour and metal insulator transition in this sample when grain size is reduced sharply in contrast to the behaviour of its bulk counterpart. We have attributed these phenomena to the destabilization of polaronic transport due to enhanced surface disorder in the nano grains. We find supportive evidence of suppression of electrical nonlinearity in the nanometric sample. The insulating, metallic and low temperature resistivity minima parts are separately fitted with conventional models applicable to each regime. A very high value of magnetoresistance (nearly 90%) is observed in all the samples at a maximum applied field of 8 T. The MR vs. H curves could not be fitted to any general power law over the whole temperature regime. It is found that spin dependent hopping model for magnetotransport is not adequate to describe the MR vs. H curves at all temperatures for all the samples. However, the spin polarized tunnelling model is found to fit well to the experimental data in the two lowest grain size samples. This implies that the magneto - transport mechanism in the samples is very complicated in deed and not unique for all the samples.

**Acknowledgement**

One of the authors (T. K. Nath) would like to acknowledge the financial assistance of Department of Science and Technology (DST), New Delhi, India through project no. IR/S2/PU-04/2006.

## Figure captions

Fig. 1. (a) Experimental x-ray diffraction data, fitted curve after Rietveld refinement and difference plot. (b) TEM image of the grains of S750 sample. (c) Size distribution of grain size, (d) SAD pattern and (e) High resolution lattice image of S750 sample.

Fig. 2. FESEM images of (a) S850, (b) S950 and (c) S1150 sample showing grain morphology. (d) EDS spectra of S750 sample showing the relative proportion of its chemical constituents.

Fig. 3. Plot of resistivity of (a) S1150, (b) S950, (c) S850 and (d) S750 samples as a function of temperature at different magnetic fields. Insets show the MR% vs. temperature plot of the corresponding samples at 8 T field.

Fig. 4. Plot of real part of ac susceptibility ($\chi_{ac}$) at 333.3 Hz an 2 Oe ac magnetic field as a function of temperature of all the samples. Inset shows the M - H behaviour of the samples at a temperature of 80 K.

Fig. 5. (a) The fitted curves (lines) with the experimental data of S1150 and S750 according to SPH model. Inset of (a) shows the nonlinear resistivity ($R_{3\omega}$) of S1150 and S750 samples. (b) Fitted curves (lines) with the experimental data in the metallic regime of the S750, S850 and S950 samples. Inset of (b) shows the fitted resistivity data in the low temperature regime

of S750 sample. (c) Resistivity in the low temperature regime of S750 at different magnetic fields. Inset of (c) shows the plot of $T_{min}$ and depth of the minima as a function of field for S750 sample. (d) Schematic presentation of the phenomenological core-shell type model where the curly arrow shows the possible path of an itinerant electron through the superimposed surfaces having weak polaronic order.

Fig. 6. Variation of MR with magnetic field at different temperatures for (a) S750, (b) S850, (c) S950 and (d) S1150 samples. The lines in (a) and (b) presents the fitted curves under the scheme of spin polarized tunnelling model.

Fig. 7. (a) Fitted curves (lines) with the experimental data at selected temperatures according to the spin dependent hopping model. Panel (b), (c), (d) and (e) show the temperature dependent phase fractions and cluster size as obtained from spin dependent hopping model for S950 and S1150 samples. (f) Temperature variation of $MR_{SPT}$ and $MR_{INT}$ of S750 and S850 samples calculated at 8 T magnetic field employing spin polarized tunnelling model.

**Table 1**

Various physical parameters like lattice parameters, unit cell volume, activation energy, and density of states at the Fermi level of the samples.

| samples | a (Å) | b (Å) | C (Å) | V ($Å^3$) | W (meV) | $\rho_0$ (ohm-cm) | $N(E_F)$ ($\times 10^{22}$) |
|---|---|---|---|---|---|---|---|
| S1150 | 5.4782 | 7.7072 | 5.4574 | 230.4 | 3 (below 35 K)<br>7 (35 K to 90 K)<br>84 (90 K to 140 K)<br>144 (above 140 K) | 3677<br>513<br>0.0023<br>0.000005 | --- |
| S950 | 5.4820 | 7.7157 | 5.4642 | 231.1 | 79 | ---- | 10.8 |
| S850 | 5.4625 | 7.6963 | 5.4528 | 229.2 | 148 | ---- | 2.3 |
| S750 | 5.4658 | 7.6982 | 5.4545 | 229.5 | 154 | --- | 2.6 |

**Table 2**

The calculated resistivity coefficients in the metallic regime and Coulomb blockade energy of the samples.

| samples | $\rho_2$(ohm-cm-$K^{-2}$) | $\rho_{4.5}$(ohm-cm-$K^{-4.5}$) | $E_C$ (K) |
|---|---|---|---|
| S950 | 3.38 | $1*10^{-18}$ | 0.01 |
| S850 | 1.71 | $1.03*10^{-6}$ | 1.21 |
| S750 | 0.15 | $5.67*10^{-7}$ | 7.8 |

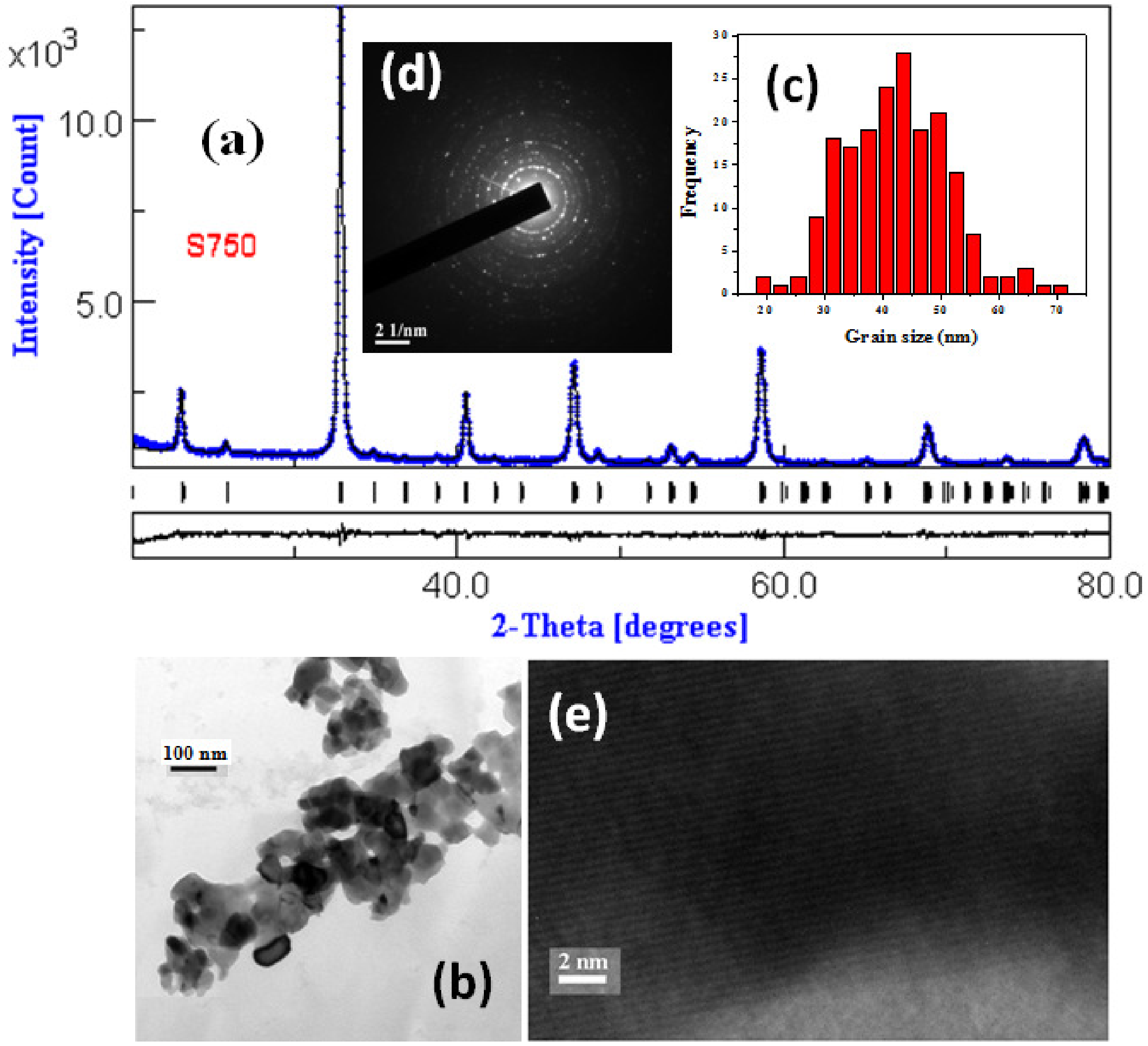

**Figure 1: S. Kundu et al.**

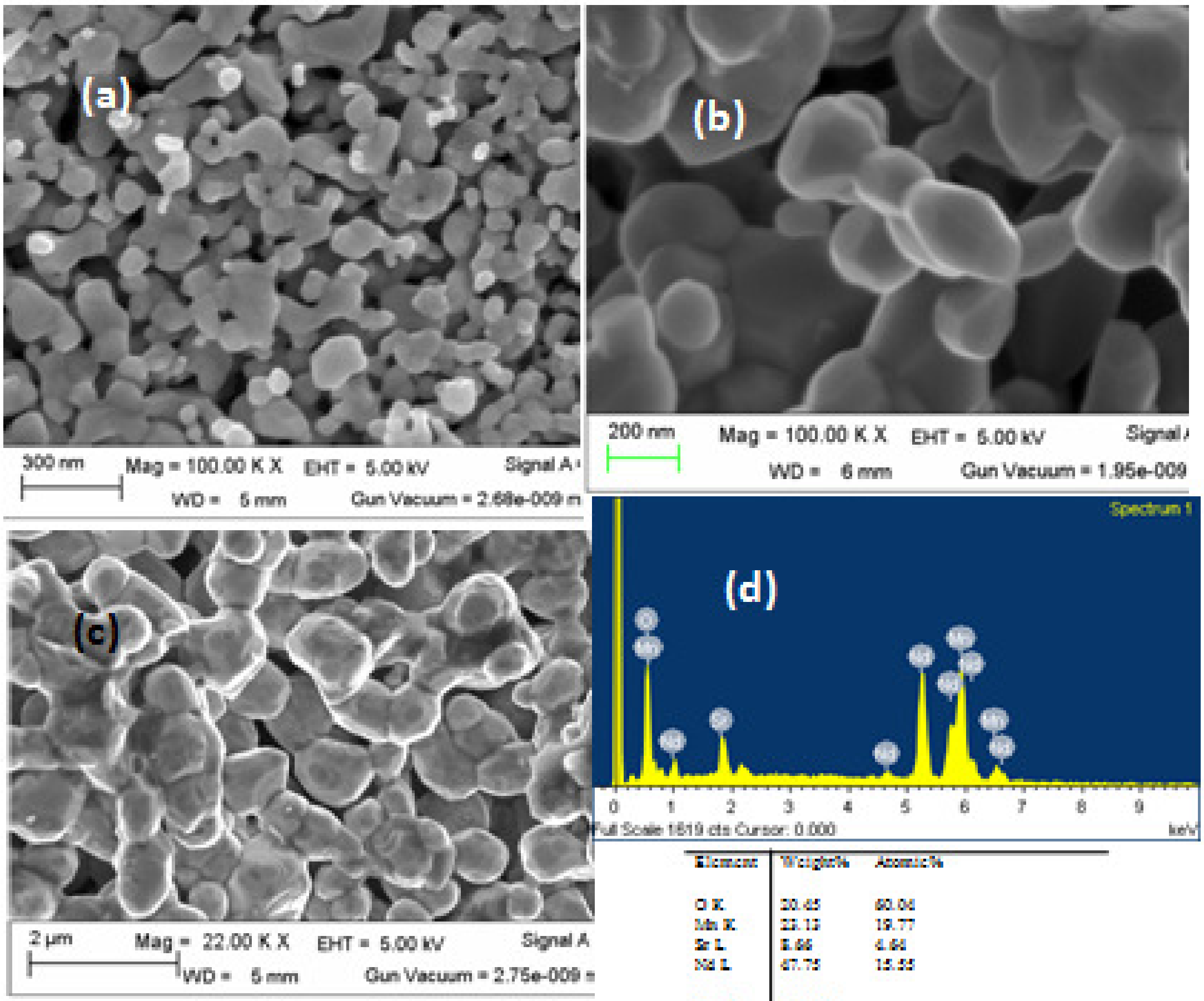

| Element | Weight% | Atomic% |
|---|---|---|
| O K | 20.45 | 60.04 |
| Mn K | 23.13 | 19.77 |
| Sr L | 8.66 | 4.66 |
| Nd L | 47.75 | 15.53 |
| Totals | 100.00 | |

**Figure 2: S. Kundu et al.**

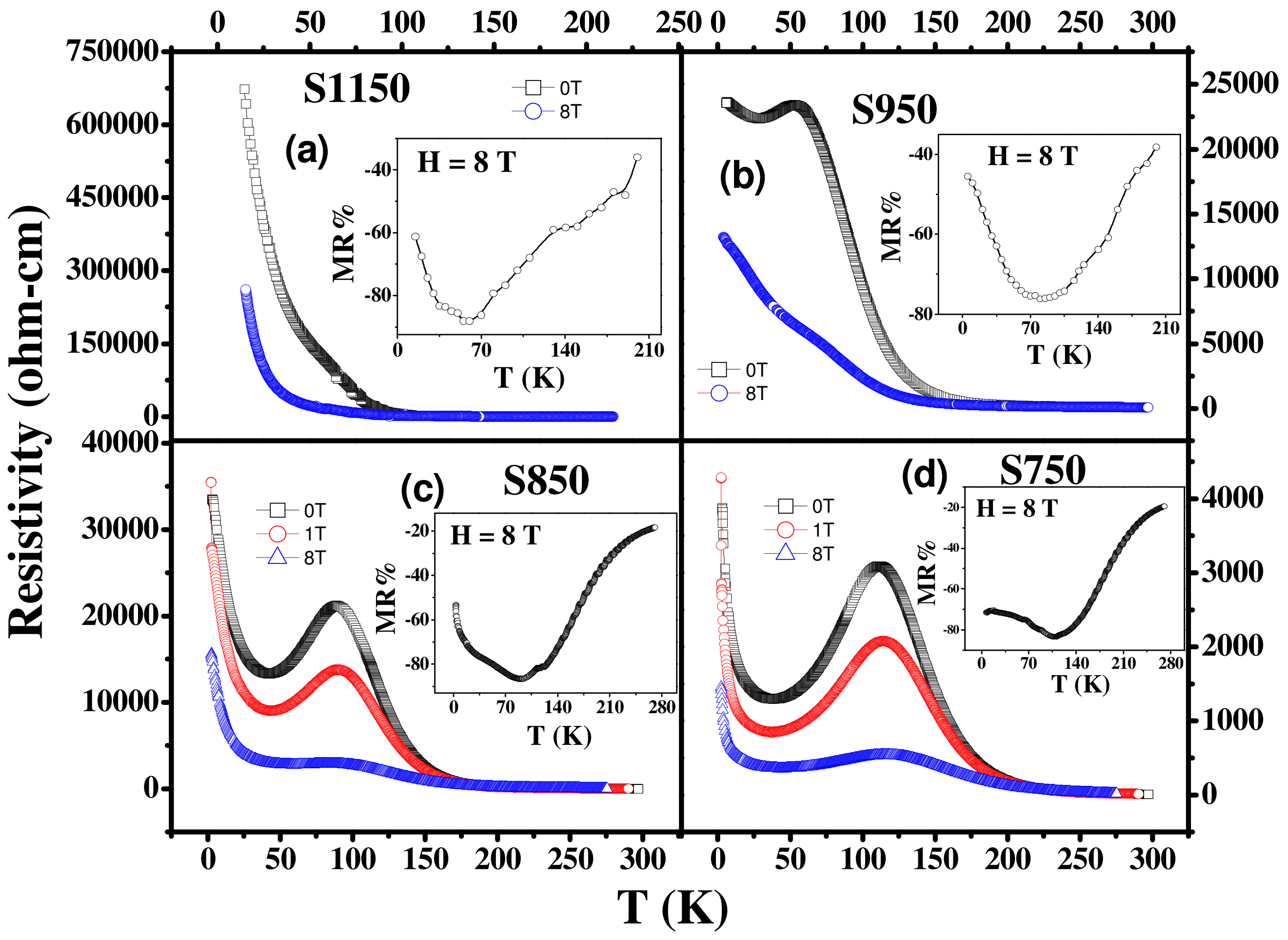

**Figure 3: S. Kundu et al.**

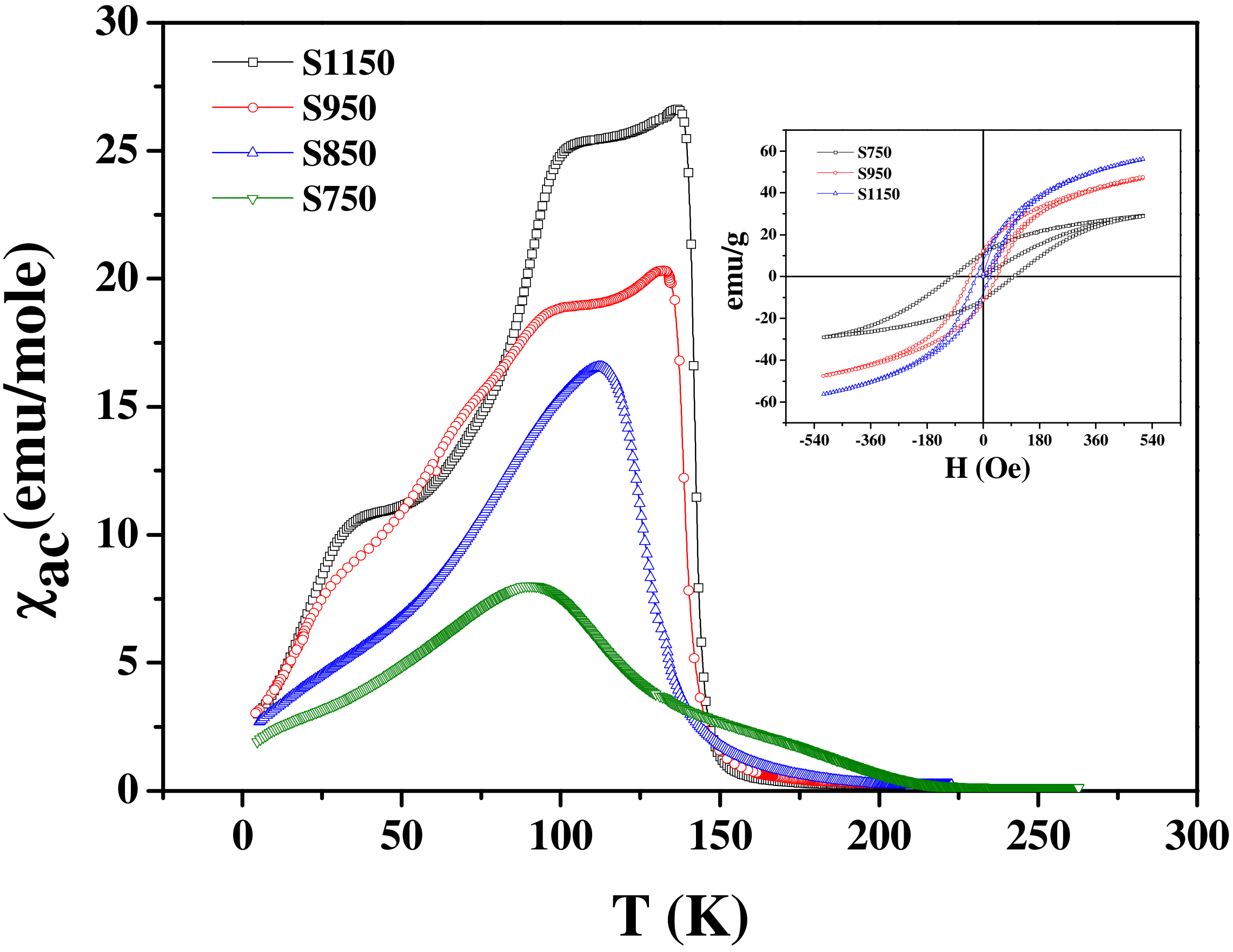

**Figure 4: S. Kundu et al.**

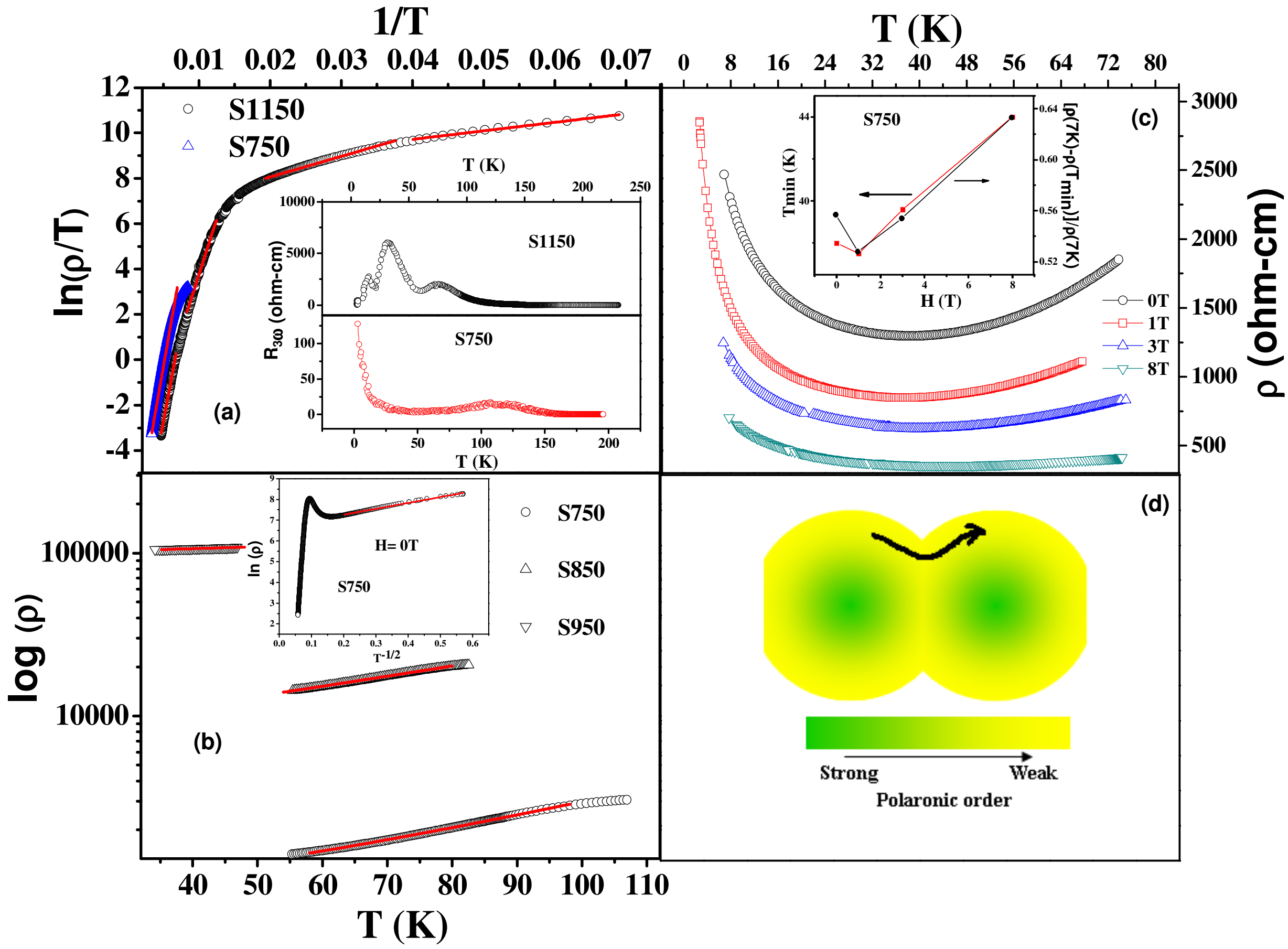

# Figure 5: S. Kundu et al.

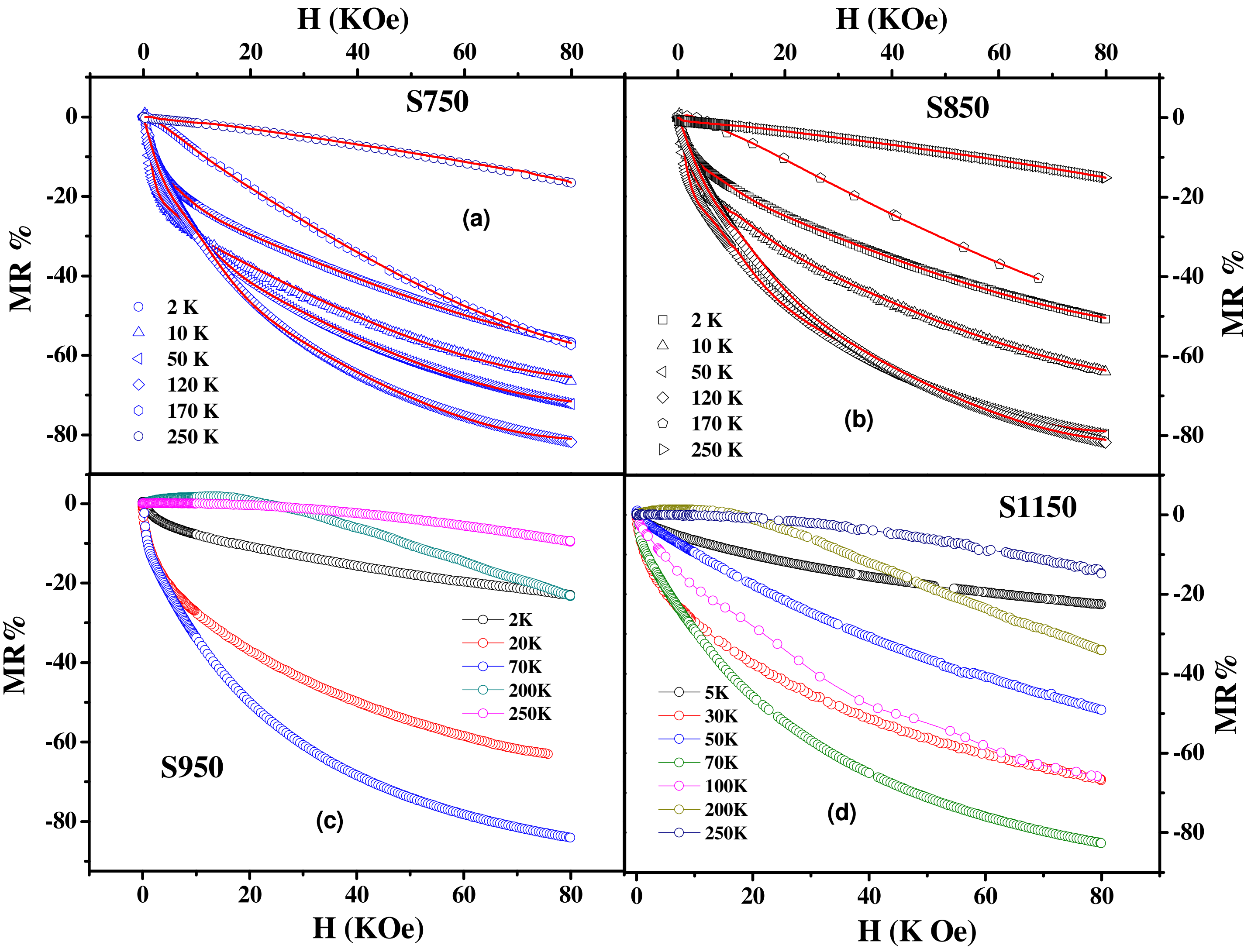

**Figure 6: S. Kundu et al.**

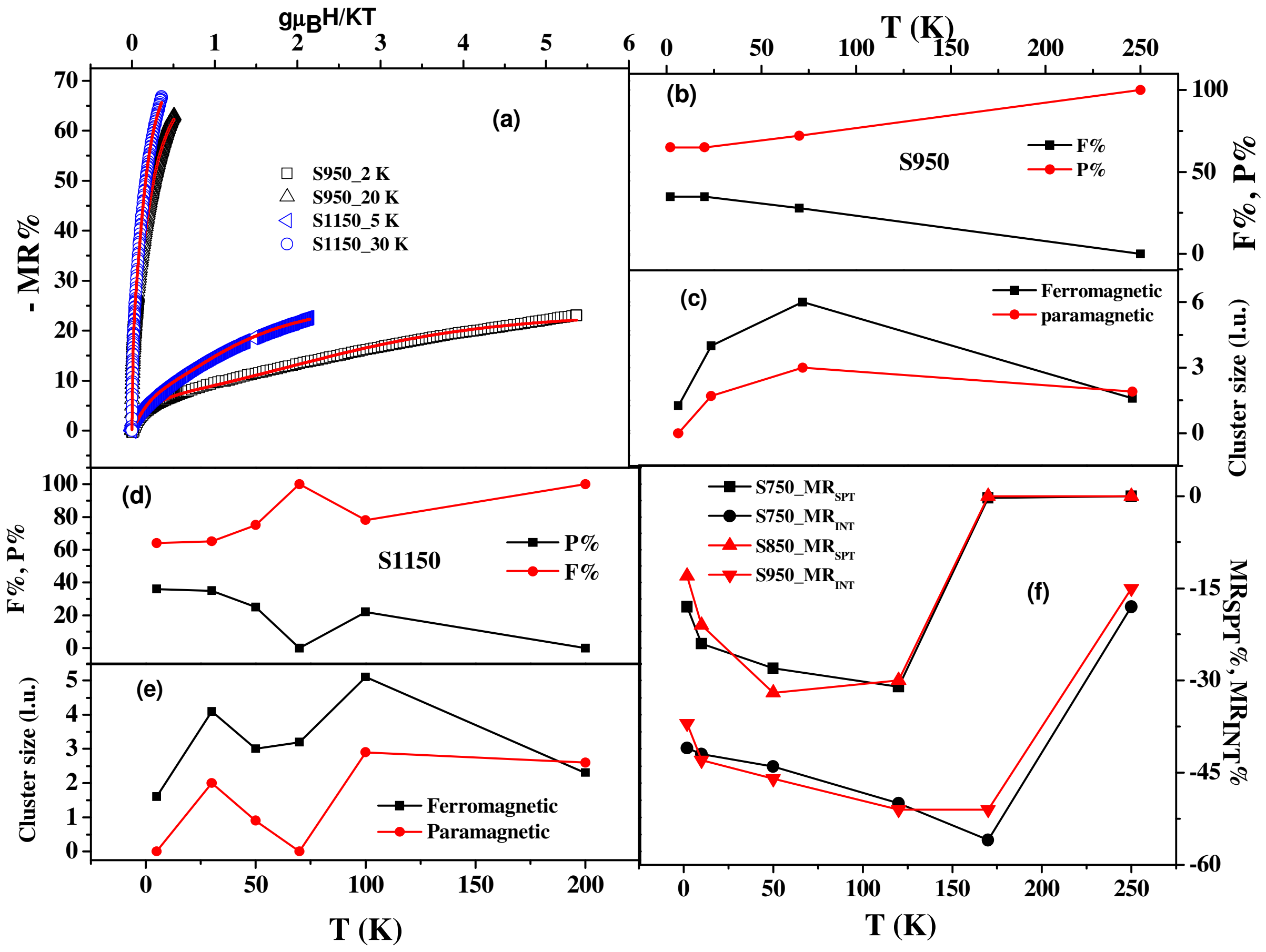

**Figure 7: S. Kundu et al.**